\documentclass[conference]{IEEEtran}
\IEEEoverridecommandlockouts
\usepackage{booktabs}
\usepackage{cite}
\usepackage{amsmath,amssymb,amsfonts}
\usepackage{algorithmic}
\usepackage{graphicx}
\usepackage{textcomp}
\usepackage{xcolor}
\usepackage{hyperref}
\newcommand{\figref}[2][{}]{\hyperref[#2]{\figurename~\ref{#2}#1}}
\usepackage{cleveref}
\usepackage{xspace}
\newcommand{\andesgym}{\texttt{andes\_gym}\xspace}
\newcommand{\andes}{\textsc{andes}\xspace}

\def\BibTeX{{\rm B\kern-.05em{\sc i\kern-.025em b}\kern-.08em
    T\kern-.1667em\lower.7ex\hbox{E}\kern-.125emX}}
\begin{document}

\title{\texttt{Andes\_gym}: A Versatile Environment for Deep Reinforcement Learning in Power Systems
}

\author{\IEEEauthorblockN{Hantao Cui}
\IEEEauthorblockA{\textit{School of Electrical and Computer Engineering} \\
\textit{Oklahoma State University}\\
Stillwater, OK USA \\
h.cui@okstate.edu}
\and
\IEEEauthorblockN{Yichen Zhang}
\IEEEauthorblockA{\textit{Energy Systems Division} \\
\textit{Argonne National Laboratory}\\
Lemont, IL USA \\
yichen.zhang@anl.gov}
}

\maketitle

\begin{abstract}
This paper presents \andesgym, a versatile and high-performance reinforcement learning environment for power system studies.
The environment leverages the modeling and simulation capability of \andes and the reinforcement learning (RL) environment OpenAI Gym to enable the prototyping and demonstration of RL algorithms for power systems.
The architecture of the proposed software tool is elaborated to provide the ``observation" and ``action" 
interfaces for RL algorithms.
An example is shown to rapidly prototype a load-frequency control algorithm based on RL trained by available algorithms.
The proposed environment is highly generalized by supporting all the power system dynamic models available in \andes and numerous RL algorithms available for OpenAI Gym.
\end{abstract}

\begin{IEEEkeywords}
Reinforcement learning, training environments, power system simulation, control
\end{IEEEkeywords}

\section{Introduction}
Reinforcement learning (RL) is a machine learning technique that enables agents to learn by trial and error in an interactive environment.
Deep reinforcement learning (DRL) combines reinforcement learning with the nonlinear feature of deep neural networks to learn from unstructured input data.
DRL has recently been applied to power system operations and controls due to its capability to accept a large volume of data for complex systems, such as the power grid.

Recent literature has explored DRL for decision-making in power systems.
Related research can be categorized by the field of application or the DRL techniques.
In the field of power system operations, DRL is applied for steady-state problems such as economic dispatch in transmission systems \cite{duan_deep_2019}, including combined heat and power plants\cite{ZHOU2020106016} and virtual power plants \cite{lin_deep_2020}, as well as in microgrids \cite{masburah_adaptive_2021}.
In power system control and dynamics, DRL is applied to transmission systems for emergency control \cite{huang_adaptive_2019}, load frequency control \cite{yan_multi-agent_2020}, frequency response model calibration \cite{li_drl-based_2021}, and voltage control \cite{thayer_deep_2020}. 
Applications are also seen in distributions systems such as Volt-Var control \cite{sun_two-stage_2021}.
This list is non-exhaustive in this emerging field.

In terms of DRL algorithms, conventional methods such as deep Q-learning \cite{yin_design_2018} and recent methods such as Proximal Policy Optimization (PPO) \cite{zhou_deriving_2020} and Deep Deterministic Policy Gradient (DDPG) \cite{zhang_economical_2020} have been applied.
Stillarious DRL algorithms are being proposed in the active ML community.
From the power engineering perspective, a versatile DRL environment is urgently needed to reduce the efforts for test driving DRL algorithms for power systems problems.

Existing work has developed and employed \textit{ad hoc} setup for DRL by combining domain software for power systems and DRL algorithms. For example, \cite{huang_adaptive_2019} developed a Gym environment that bridges Java-based InterPSS with Python-based DRL environment.
In \cite{zhang_hybrid_2021}, PSS/E is integrated with a unit commitment program through the \texttt{psspy} interface. In \cite{beyer_adaptive_2021}, MATLAB/Simulink is connected with OpenAI Gym for training.
In addition, generalized environments are proposed, such as the OpenModelica-Microgrid-Gym environment in \cite{heid_omg_2020}.
While the work in \cite{heid_omg_2020} is conceptually similar to this paper, our environment is fully implemented in Python and can be readily integrated with DRL frameworks, most of which provide a Python interface.

In this paper, a DRL environment is proposed by leveraging \andes \cite{cui_hybrid_2021} for power system simulation and OpenAI Gym \cite{brockman_openai_2016} to interface to various algorithms.
The proposed work, \andesgym is versatile for prototyping controls for power flow and transient dynamics supported by \andes, which is an open-source, generalized power system simulation tool supporting industry-grade models and PSS/E input files. \andesgym is maximal in performance given the ahead-of-time compilation of numerical code in \andes and the efficient DRL training algorithms interfaced through OpenAI Gym.
\andesgym is open-source and available on GitHub 
\footnote{Repository link: \href{https://github.com/sensl/andes\_gym}{https://github.com/sensl/andes\_gym}}.

The paper is organized as follows: \Cref{sec:architecture} introduces the architecture and workflow of the proposed environment and used load-frequency control as an example, \Cref{sec:case-studies} presents the training results of a load-frequency controller in the proposed environment, using the IEEE 14-bus system and the DDPG algorithm.
\Cref{sec:conclusion} draws the conclusions. 

\section{Methodology and Software Architecture}
\label{sec:architecture}
The key components to training RL agents are the environment representing the system being observed and controlled, as well as the RL algorithms that process the observations and generate the actions.
The environment will provide key functions such as advancing the time, simulating system dynamics, taking observations, accepting action signals, and starting over for a new training episode.
This section will discuss the integration of \andes with OpenAI Gym, demonstrated by an environment to train a DRL-based load frequency controller.

\subsection{Architecture and Components}

\begin{figure}[t]
\centerline{\includegraphics[width=\columnwidth]{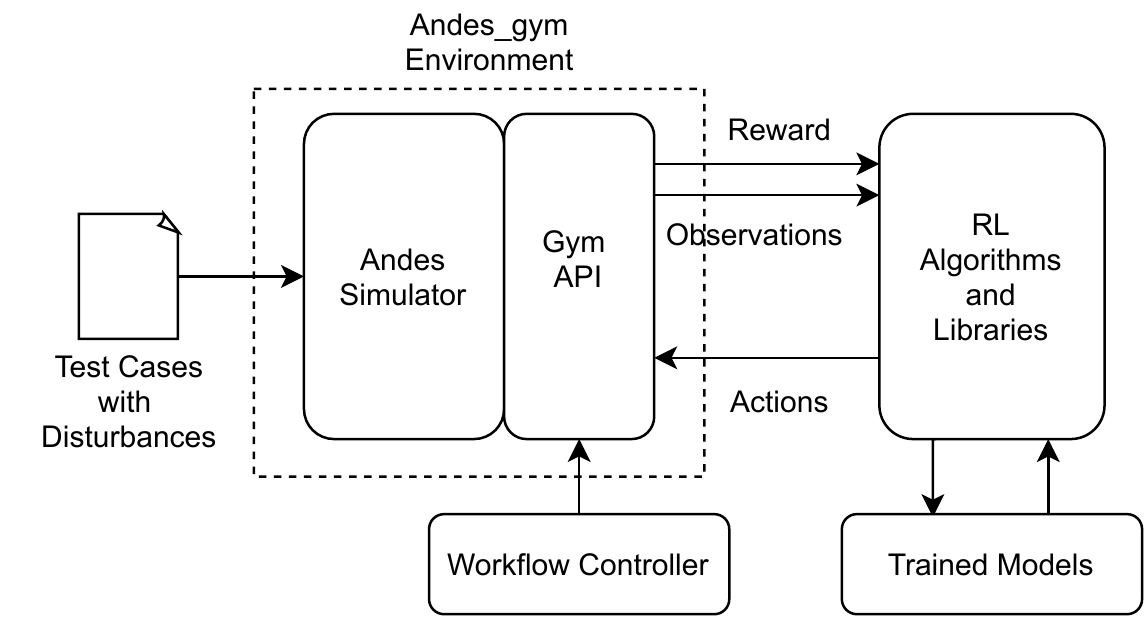}}
\caption{Architecture of the \andesgym environment for RL.}
\label{fig:architecture}
\end{figure}

\figref{fig:architecture} shows the general architecture and components of an RL environment for power system studies.
The main components are the \andes simulator, the OpenAI Gym application programming interface (API) implemented as port of the environment, and reinforcement learning algorithms.
Details of the components are described as follows.

In a power system RL environment, it is natural to represent the system dynamics with simulation software that can provide models, simulation methods, and workflow control.
The \andesgym project utilizes the fully open-source \andes simulation, and the advantages are elaborated in the next subsection.
The \andesgym environment wraps \andes to provide the APIs in the OpenAI Gym specification.
Gym is a widely-used general-purpose toolkit for developing and benchmarking RL algorithms.
Gym provides abstractions for the environment to maximize the convenience to implement agents of various styles.
Although Gym is environment-only, numerous state-of-the-art open-source libraries with extensively tested algorithms, such as OpenAI Baseline, have been interfaced to Gym and are ready for training.

When setting up, the user specifies in the environment the appropriate test case with desired disturbances.
Multiple environments can be created and tailored for specific applications by running systems representing various dynamic behaviors, measuring different states, and actuating with different devices.

It is worth noting that the data interfaces in the architecture are highly efficient. 
Between any two components, data interfacing is performed in memory, which minimizes the overhead of inputs and outputs (I/Os).
This is an advantage of having the simulation program, the environment, and the training algorithms available in the same programming language.
File I/Os only happen when \andes reads the power system test cases and when the trained RL agents are stored.

\subsection{Modeling and Simulation with \andes}
Although various tools exist for simulation, \andes is used for this work due to the following advantages:

\begin{itemize}
    \item \andes is fully open-source and written in Python, which can be readily picked up by researchers and industry users for rapid prototyping.
    \item \andes has a descriptive modeling framework that allows advanced users to quickly prototype new models and controllers using equations and modeling blocks. Such blocks include the commonly used transfer functions and discrete components.
    The built-in processing and code generation functions will convert the descriptive models into executable code.
    It is especially useful when one needs to test and control a new model that has not existed in commercial software.
    \item Based on the descriptive modeling framework, \andes provides the full implementation of commercial-grade models, including generator models and controllers, as well as the second-generation renewable models. 
    \item \andes has recently adopted \texttt{numba} for compilation. It compiles the generated code for model equations and Jacobian functions
    into machine code to achieve near-native performance.
    \andes enables the compilation ahead of time to minimize any runtime overhead.
    \item \andes is easy to use. It comes with a built-in parser for PSS/E \texttt{raw} format for power flow and \texttt{dyr} format for dynamics, which is enabled by the identical dynamic models implemented in \andes.
\end{itemize}

\subsection{RL Workflow Control for Power System Simulation}

The OpenAI Gym interface defines the necessary steps for RL, including test case construction, initialization, stepping, resetting, and optionally, rendering.
The \andesgym environment implements the following interfaces:
\begin{itemize}
    \item \texttt{\_\_init\_\_()}: this is the constructor of the environment for setting up hyper-parameters.
    Such parameters include the test case, simulation parameters, the instants at which the system is observed and actions are taken are specified, and the limits of the actions. 
    Parameters provided in this interface are consider static.
    \item \texttt{initialize()}: this is the run-time initialization function for setting up the environment.
    For power system simulations, power flow solution and the initialization of time-domain simulations are performed.
    \item \texttt{step()}: this is the stepping function for controlling the \andes simulation. 
    It accepts an action generated from previous observations.
    In our design, the stepping function will progress the simulation time to the next observation and action point as defined in the parameter set. 
    The reward formula, which is crucial for RL problems, is implemented in this function.
    At the end, it returns the current observation, the cumulative rewards, and a flag to indicate whether or not the simulation has ended.
    \item \texttt{reset()}: it resets the system for the next episode. 
    \item \texttt{seed()}:  it is the interface for seeding the random number generator, which allows creating repeatable pseudo-random scenarios.
\end{itemize}

\begin{figure}[t]
\centerline{\includegraphics[width=0.7\columnwidth]{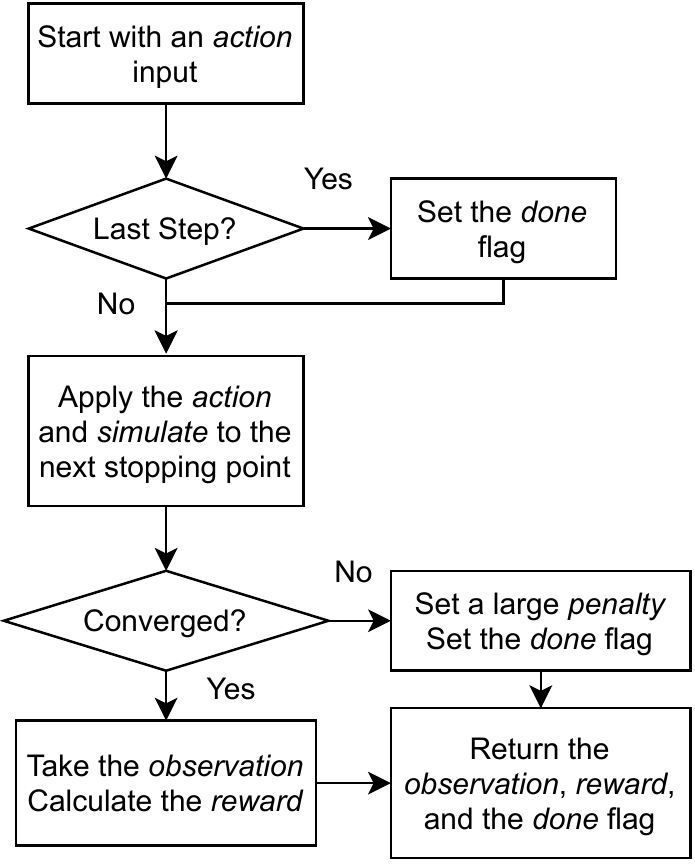}}
\caption{Flowchart of the stepping function.}
\label{fig:step}
\end{figure}

\figref{fig:step} shows the workflow in the stepping interface. It needs to be distinguished that the step in the RL workflow is different from the step size in the simulation program.
Typically, the step size in time-domain simulation is smaller than the RL, meaning that each step in the RL workflow wraps multiple steps in the simulation before observations are obtained and actions are taken.
This is relevant when the simulation granularity is much greater than RL-based controls.

\subsection{\andesgym for Load Frequency Control}
An example environment is created for load frequency control in power systems.
The environment is set up to include load variations, and
RL agents will need to control the reference signals for turbine governors to bring the frequency back to nominal.
Fundamentally, the RL agents act as a secondary frequency controller but at a higher rate.
That is, unlike the traditional automatic generation control (AGC) that updates the reference signals at four-second intervals, the RL-based frequency controller modified the power set point at every time instant that the system is observed.

\begin{figure}[t]
\centerline{\includegraphics[width=\columnwidth]{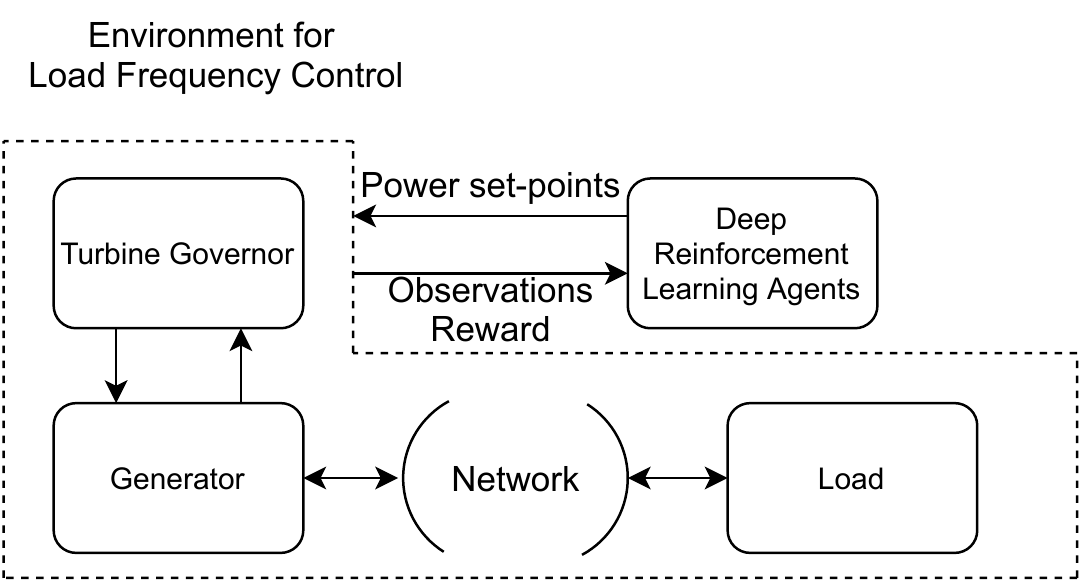}}
\caption{Flowchart of the stepping function.}
\label{fig:drl-outline}
\end{figure}

\figref{fig:drl-outline} shows the main components for DRL-based frequency controller using the proposed environment.
In this scenario, the speed of generators is observed from the grid, although additional information such as the rate-of-change of frequency (RoCoF) can be used. 

\section{Case Studies}
\label{sec:case-studies}
\subsection{IEEE 14-Bus System}
\andesgym is utilized to train the DRL agent for load frequency control.
The IEEE 14-bus system is utilized with a load change event of 60 MW (or 0.1 pu in the system base of 100 MVA) on Bus 4 at 1 s.
All the loads are represented by constant power models.
There are five generators in the system, and all their rotor speed is observed.
All the five turbine governors connected to the five generators are controlled with the action space for the DRL agent set to $[-10, 10]$ MW.
The scenarios will study the impacts of delay and the number of actions on the training convergence and control performance. 

\subsection{Learning Problem Setup}

\begin{figure}[t]
\centerline{\includegraphics[width=0.9\columnwidth]{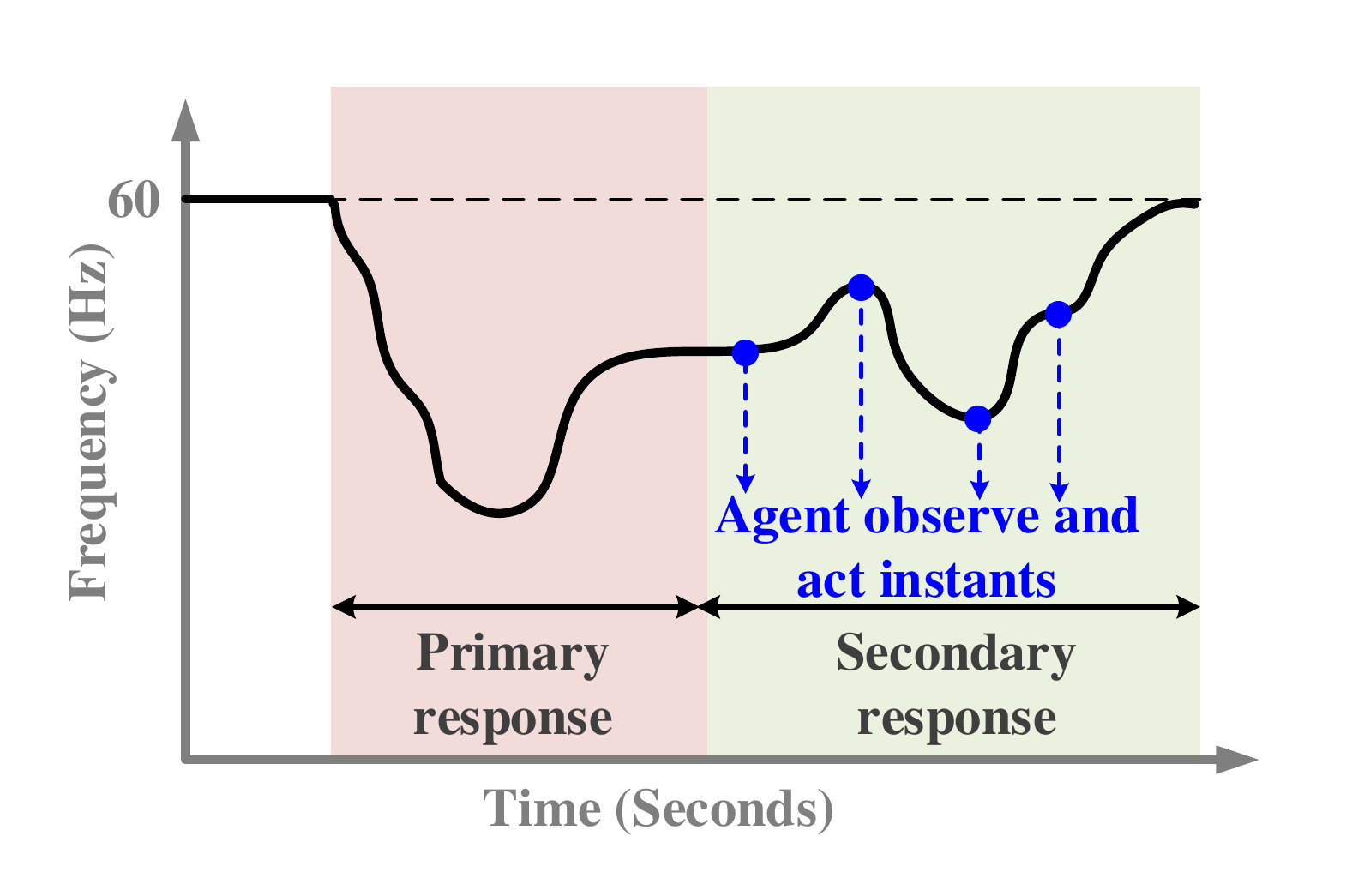}}
\caption{The observation and action instants of the agent.}
\label{fig:drl-traning_setup}
\end{figure}

The most important information for secondary frequency control is the post-disturbance steady-state frequency, which is directly associated with the value of the disturbance. Essentially, the goal is to encode the disturbance estimator with the control policy to build an end-to-end one. Therefore, the agent begins to observe and act when the frequency comes to a steady state. Then, the agent can observe and act consecutively at designated instants, which is denoted as the \emph{action instants},  as shown in Fig. \ref{fig:drl-traning_setup}. The interval between the action instants is predefined as one of the hyper-parameters for the DRL problem. Consider the fact that the action instants in the middle is to let the agent learn the frequency dynamic behaviors under control inputs while at the last instant the frequency should be restored to the nominal condition, the reward function at instant $t$ is designed as follows:
\[
    r_t= 
\begin{cases}
    -100(f_t - f^{*}),& \text{if } t\neq T_{f}\\
    -3000(f_t-f^{*}),              & \text{otherwise}
\end{cases}
\]
where $f_t$ is the frequency observation and $f^{*}$ is the nominal frequency. The term $T_{f}$ is the final instant, at which the frequency deviation is penalized with a large weight. For the instants in the middle, a relatively small penalty is used. The reason is that the frequency might not be able to restore back between instants due to physical constraints. Setting an equally large weight with the final instant will prohibit the agent from exploration. The expectation is by conducting consecutive actions the agent could restore the frequency at the final instant. In this study,  the agent is allowed to observe and act between 5 and 10 seconds after the disturbance and expect that the agent can restore the frequency back to 60 Hz at the end of 10 s. 
In addition, a \texttt{learning\_start} parameter, which controls when the agent should start learning and has a major impact, is considered. The parameter controls the balance between exploration and exploitation. In the study, a fixed disturbance of 0.6 pu is applied at 1 s. The deep deterministic policy gradient (DDPG) method is employed from \texttt{stable\_baseline} to solve the problem. In the next subsection, the training performance will be compared under different settings of \texttt{learning\_start} and action intervals.

\subsection{Learning Performance}\label{AA}
\begin{figure}[t]
\centerline{\includegraphics[width=\columnwidth]{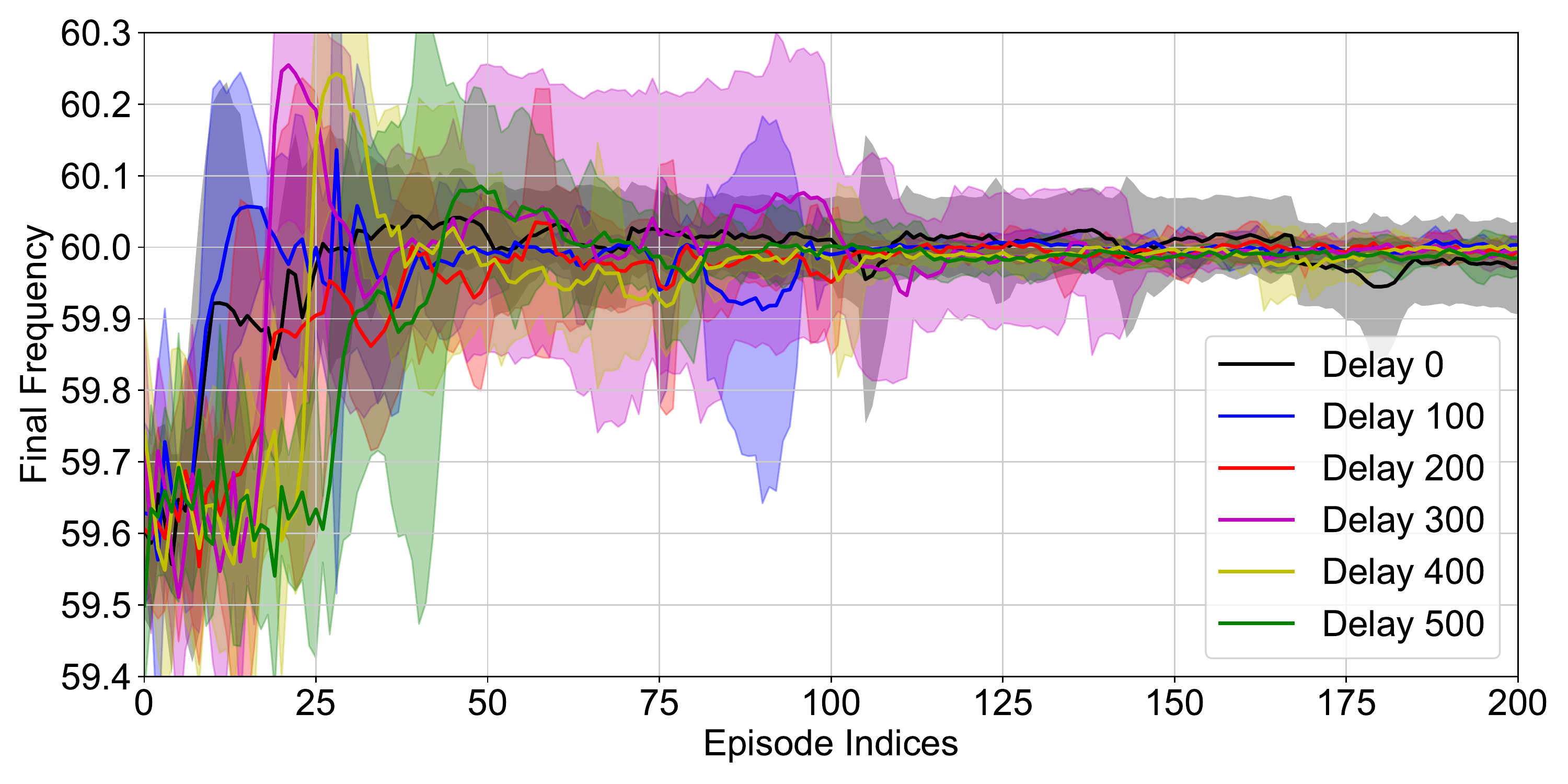}}
\caption{The final frequency values at the end of each episode under different \texttt{learning\_start} settings. The solid lines are the averaged values while the colored areas are the standard deviations.}
\label{fig:drl-delay}
\end{figure}

In the first computation experience, we fix the action number to be 20. Meanwhile, different settings of \texttt{learning\_start} are considered. 
The \texttt{learning\_start} is set to 0, 100, 200, 300, 400, and 500. 
In each episode, the frequency value at the final instant $T_f$ is recorded to measure how well the agent is trained.
Ten runs for each setting are performed.
The final frequency values at the end of each episode under different \texttt{learning\_start} settings are plotted in Fig. \ref{fig:drl-delay}. The solid lines are the averaged values while the colored areas are the standard deviations. As shown, without delayed learning, the final frequency variations between 15th and 75th episodes are relatively small. However, the variations after 100 episodes are relatively large, which is approximately equal to 0.1 Hz. Additionally, there exists a small deviation in the averaged value. When the \texttt{learning\_start} is set to be 200, the final frequency variations between 15th and 75th episodes are larger since pure random actions are employed. After the training started, the averaged frequency quickly converged to the nominal value with a small variation. A similar process is observed when the \texttt{learning\_start} is set to be 500 expect the variations are larger. 
The averaged frequency deviation values of the last 50 episodes are shown in Table \ref{tab_delay}. It shows that a delayed learning setting of 100 has slightly better performance than the other ones. 
This is expected due to the balance of efforts between exploration and exploitation to improve the overall performance. 

\begin{table}[t]
	\caption{Last 50 Episodes Averaged Frequency Deviation Values }
	\centering
	\begin{tabular}{c c}
		\hline\hline 
		Delay Learning & Frequency Deviation  \\
		\midrule
		0 & 0.0233 Hz \\
		100 & 0.0042 Hz\\
		200 & 0.0045 Hz \\
		300 & 0.0067 Hz\\
		400 & 0.0093 Hz \\
		500 & 0.0108 Hz\\
		\hline \hline 
	\end{tabular}
	\label{tab_delay}
\end{table}

\begin{figure}[t]
\centerline{\includegraphics[width=\columnwidth]{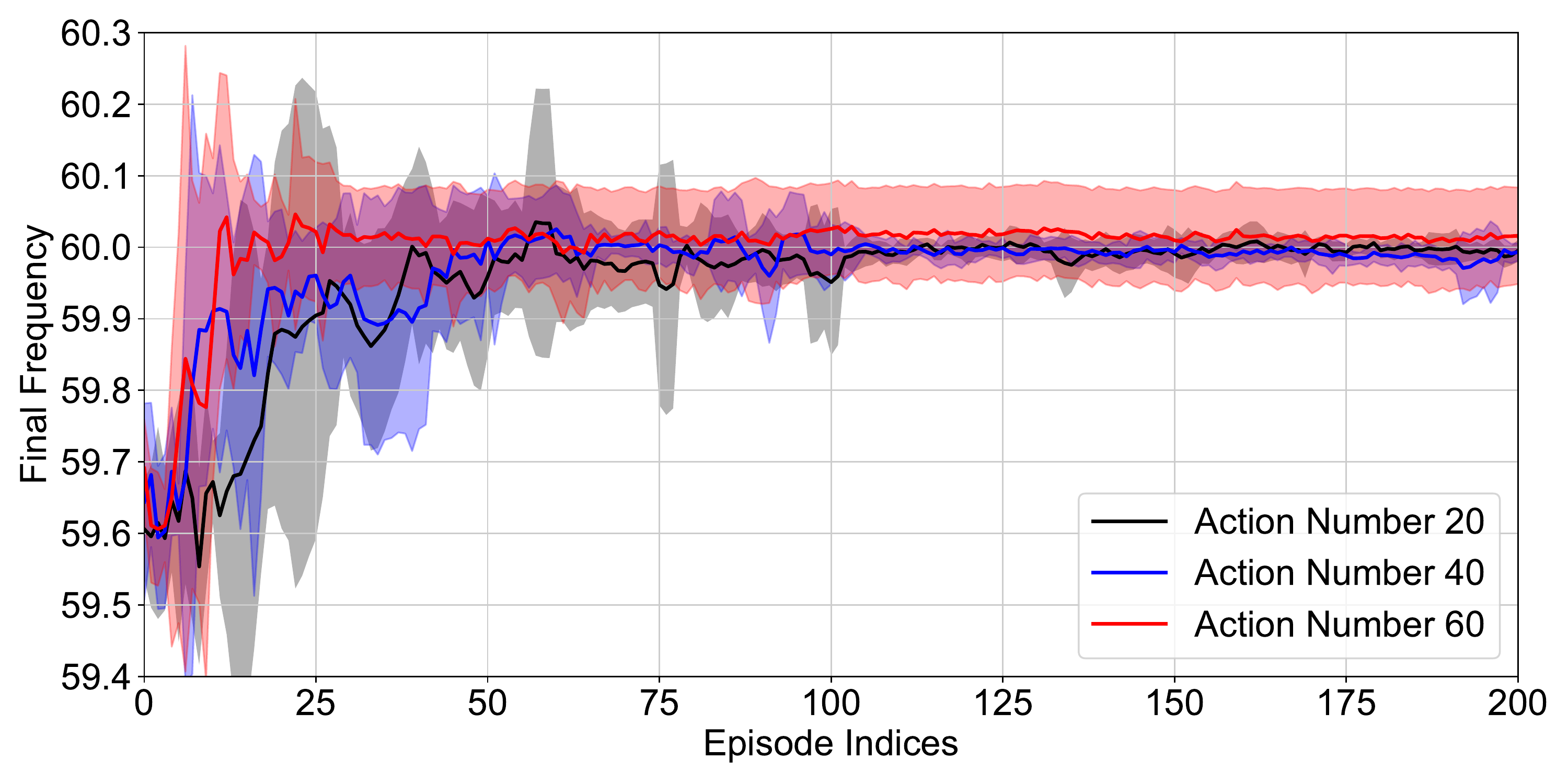}}
\caption{The final frequency values at the end of each episode with different numbers of actions. The solid lines are the averaged values while the colored areas are the standard deviations. }
\label{fig:drl-action}
\end{figure}

In the second computation experiment, different settings of action intervals are considered. It is equivalent to varying the number of actions between 5 to 10 seconds. In particular, the number of actions is set to 20, 40, and 60.
Based on the previous study, the learning delay is set to 200 steps. The same performance metrics is used, namely, the frequency value at the final instant in each episode. 
Ten runs for each setting are conducted.
The final frequency values at the end of each episode under different action numbers are plotted in Fig. \ref{fig:drl-action}. The solid lines are the averaged values while the colored areas are the standard deviations. Based on Fig. \ref{fig:drl-action}, the training performance decreases with the increase of action numbers. Before 100 episodes, all three settings are exploring to collect state action pair and associated rewards. After 100 episodes, the first two settings quickly converge to the nominal frequency with small variations, meaning both agents admit a high success rate to restore the frequency. With the action number equaling to 60, the agent's frequency restoration has a relatively large steady-state error compared with the previous two cases. Additionally, the variations are larger, indicating that the agent can fail to restore the frequency in some episodes. The averaged frequency deviation values of the last 50 episodes are shown in Table \ref{tab_delay}. The results are consistent with the plots. 

The rationale behind this is that the secondary frequency control mainly concerns the steady state. Therefore, the most informative states to the agent are the value of the steady-state frequency under a trial control input. With sufficient numbers of such trials, the agent is able to learn the appropriate control value to steer the frequency back to nominal. If the action instants are too frequent, the frequency is still under the transient period. 
In this case, additional rate-of-change of frequency information is included, and the agent will need more samples to learn the useful part for secondary control. In short, less action numbers filter out the less relevant information and only provide the essential states related to the secondary frequency control, and thus converge faster. 

\begin{table}[t]
	\caption{Last 50 Episodes Averaged Frequency Deviation Values }
	\centering
	\begin{tabular}{c c}
		\hline\hline 
		Action Number & Frequency Deviation  \\
		\midrule
		20 & 0.0046 Hz \\
		40 & 0.0109 Hz \\
		60 & 0.0136 Hz\\
		\hline \hline 
	\end{tabular}
	\label{tab:last-50}
\end{table}

\begin{figure}[htbp]
\centerline{\includegraphics[width=\columnwidth]{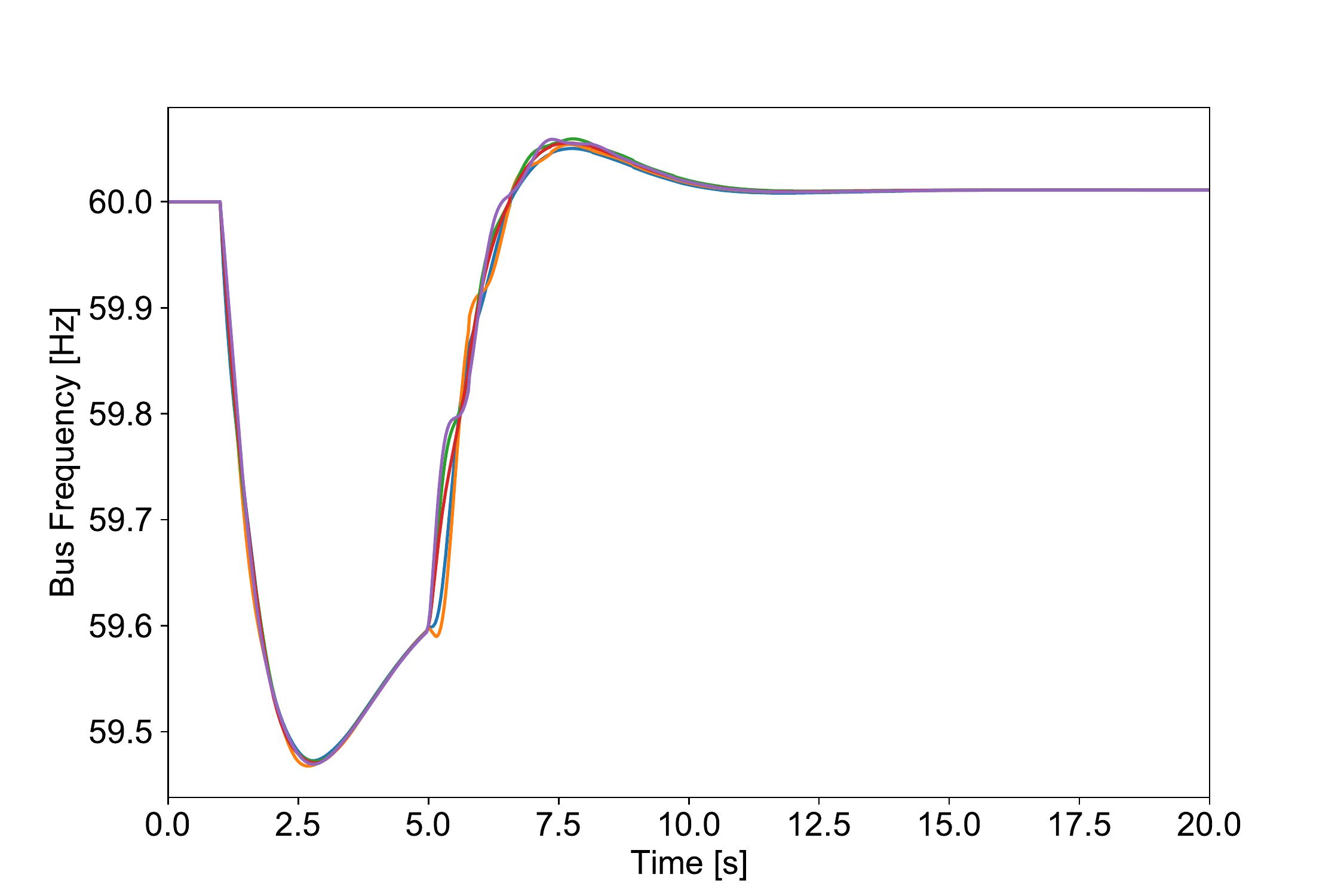}}
\caption{The frequency dynamics under the trained agent.}
\label{fig:freq_dyn}
\end{figure}

The trained agent is validated using full-order nonlinear simulation in \texttt{Andes}. A 0.6 per unit disturbance is applied at 1 second. The agent began to observe and act at 5 seconds. The simulation result is shown in Fig. \ref{fig:freq_dyn}. As shown, the agent can successfully steer the frequency back to 60 Hz.

\section{Conclusion}
\label{sec:conclusion}
This paper introduces \andesgym, a versatile and high-performance reinforcement learning environment for power system studies.
The environment leverages the modeling and simulation capability of \andes and the reinforcement learning (RL) environment OpenAI Gym to enable the development and validation of RL algorithms for power systems.
The structure, workflow, and a case study of secondary frequency control are expressed. 
\texttt{Andes\_gym} can be used to model various power system problems. 
With the standardized \texttt{Gym} I/Os, the created problems can be solved using either widely used reinforcement learning solvers, such as \texttt{stable\_baseline}, or customized RL algorithms.
This feature also allows fast comparisons between multiple RL algorithms. 
Case studies also demonstrate the convenience to compare the impacts of hyper-parameter on the training performance using \texttt{andes\_gym}.

\bibliographystyle{IEEEtran}
\bibliography{IEEEabrv,papers}

\begin{thebibliography}{10}
\providecommand{\url}[1]{#1}
\csname url@samestyle\endcsname
\providecommand{\newblock}{\relax}
\providecommand{\bibinfo}[2]{#2}
\providecommand{\BIBentrySTDinterwordspacing}{\spaceskip=0pt\relax}
\providecommand{\BIBentryALTinterwordstretchfactor}{4}
\providecommand{\BIBentryALTinterwordspacing}{\spaceskip=\fontdimen2\font plus
\BIBentryALTinterwordstretchfactor\fontdimen3\font minus
  \fontdimen4\font\relax}
\providecommand{\BIBforeignlanguage}[2]{{%
\expandafter\ifx\csname l@#1\endcsname\relax
\typeout{** WARNING: IEEEtran.bst: No hyphenation pattern has been}%
\typeout{** loaded for the language `#1'. Using the pattern for}%
\typeout{** the default language instead.}%
\else
\language=\csname l@#1\endcsname
\fi
#2}}
\providecommand{\BIBdecl}{\relax}
\BIBdecl

\bibitem{duan_deep_2019}
J.~Duan, H.~Li, X.~Zhang, R.~Diao, B.~Zhang, D.~Shi, X.~Lu, Z.~Wang, and
  S.~Wang, ``A deep reinforcement learning based approach for optimal active
  power dispatch,'' in \emph{2019 {IEEE} {Sustainable} {Power} and {Energy}
  {Conference} ({iSPEC})}.\hskip 1em plus 0.5em minus 0.4em\relax IEEE, 2019,
  pp. 263--267.

\bibitem{ZHOU2020106016}
\BIBentryALTinterwordspacing
S.~Zhou, Z.~Hu, W.~Gu, M.~Jiang, M.~Chen, Q.~Hong, and C.~Booth, ``Combined
  heat and power system intelligent economic dispatch: A deep reinforcement
  learning approach,'' \emph{International Journal of Electrical Power \&
  Energy Systems}, vol. 120, p. 106016, 2020. [Online]. Available:
  \url{https://www.sciencedirect.com/science/article/pii/S0142061519336713}
\BIBentrySTDinterwordspacing

\bibitem{lin_deep_2020}
L.~Lin, X.~Guan, Y.~Peng, N.~Wang, S.~Maharjan, and T.~Ohtsuki, ``Deep
  reinforcement learning for economic dispatch of virtual power plant in
  internet of energy,'' \emph{IEEE Internet of Things Journal}, vol.~7, no.~7,
  pp. 6288--6301, 2020, publisher: IEEE.

\bibitem{masburah_adaptive_2021}
R.~Masburah, R.~L. Jana, A.~Khan, S.~Xu, S.~Lan, S.~Dey, and Q.~Zhu, ``Adaptive
  {Learning} {Based} {Building} {Load} {Prediction} for {Microgrid} {Economic}
  {Dispatch},'' in \emph{2021 {Design}, {Automation} \& {Test} in {Europe}
  {Conference} \& {Exhibition} ({DATE})}.\hskip 1em plus 0.5em minus
  0.4em\relax IEEE, 2021, pp. 72--75.

\bibitem{huang_adaptive_2019}
Q.~Huang, R.~Huang, W.~Hao, J.~Tan, R.~Fan, and Z.~Huang, ``Adaptive power
  system emergency control using deep reinforcement learning,'' \emph{IEEE
  Transactions on Smart Grid}, vol.~11, no.~2, pp. 1171--1182, 2019, publisher:
  IEEE.

\bibitem{yan_multi-agent_2020}
Z.~Yan and Y.~Xu, ``A multi-agent deep reinforcement learning method for
  cooperative load frequency control of a multi-area power system,'' \emph{IEEE
  Transactions on Power Systems}, vol.~35, no.~6, pp. 4599--4608, 2020,
  publisher: IEEE.

\bibitem{li_drl-based_2021}
H.~Li, Y.~Xiang, T.~Jin, S.~Wang, Z.~Yi, and D.~Shi, ``A {DRL}-{Based}
  {Approach} for {System} {Frequency} {Response} {Model} {Calibration},'' in
  \emph{2020 52nd {North} {American} {Power} {Symposium} ({NAPS})}.\hskip 1em
  plus 0.5em minus 0.4em\relax IEEE, 2021, pp. 1--6.

\bibitem{thayer_deep_2020}
B.~L. Thayer and T.~J. Overbye, ``Deep reinforcement learning for electric
  transmission voltage control,'' in \emph{2020 {IEEE} {Electric} {Power} and
  {Energy} {Conference} ({EPEC})}.\hskip 1em plus 0.5em minus 0.4em\relax IEEE,
  2020, pp. 1--8.

\bibitem{sun_two-stage_2021}
X.~Sun and J.~Qiu, ``Two-{Stage} {Volt}/{Var} {Control} in {Active}
  {Distribution} {Networks} with {Multi}-{Agent} {Deep} {Reinforcement}
  {Learning} {Method},'' \emph{IEEE Transactions on Smart Grid}, 2021,
  publisher: IEEE.

\bibitem{yin_design_2018}
L.~Yin, T.~Yu, and L.~Zhou, ``Design of a novel smart generation controller
  based on deep {Q} learning for large-scale interconnected power system,''
  \emph{Journal of Energy Engineering}, vol. 144, no.~3, p. 04018033, 2018,
  publisher: American Society of Civil Engineers.

\bibitem{zhou_deriving_2020}
Y.~Zhou, B.~Zhang, C.~Xu, T.~Lan, R.~Diao, D.~Shi, Z.~Wang, and W.-J. Lee,
  ``Deriving {AC} {OPF} {Solutions} via {Proximal} {Policy} {Optimization} for
  {Secure} and {Economic} {Grid} {Operation},'' \emph{arXiv preprint
  arXiv:2003.12584}, 2020.

\bibitem{zhang_economical_2020}
B.~Zhang, W.~Hu, D.~Cao, Q.~Huang, Z.~Chen, and F.~Blaabjerg, ``Economical
  operation strategy of an integrated energy system with wind power and power
  to gas technology–a {DRL}-based approach,'' \emph{IET Renewable Power
  Generation}, vol.~14, no.~17, pp. 3292--3299, 2020, publisher: IET.

\bibitem{zhang_hybrid_2021}
Y.~Zhang, F.~Qiu, T.~Hong, Z.~Wang, and F.~F. Li, ``Hybrid {Imitation}
  {Learning} for {Real}-{Time} {Service} {Restoration} in {Resilient}
  {Distribution} {Systems},'' \emph{IEEE Transactions on Industrial
  Informatics}, pp. 1--1, 2021, conference Name: IEEE Transactions on
  Industrial Informatics.

\bibitem{beyer_adaptive_2021}
K.~Beyer, R.~Beckmann, S.~Geißendörfer, K.~v. Maydell, and C.~Agert,
  ``Adaptive {Online}-{Learning} {Volt}-{Var} {Control} for {Smart} {Inverters}
  {Using} {Deep} {Reinforcement} {Learning},'' \emph{Energies}, vol.~14, no.~7,
  p. 1991, 2021, publisher: Multidisciplinary Digital Publishing Institute.

\bibitem{heid_omg_2020}
S.~Heid, D.~Weber, H.~Bode, E.~Hüllermeier, and O.~Wallscheid, ``{OMG}: {A}
  scalable and flexible simulation and testing environment toolbox for
  intelligent microgrid control,'' \emph{Journal of Open Source Software},
  vol.~5, no.~54, p. 2435, 2020.

\bibitem{cui_hybrid_2021}
H.~Cui, F.~Li, and K.~Tomsovic, ``Hybrid {Symbolic}-{Numeric} {Framework} for
  {Power} {System} {Modeling} and {Analysis},'' \emph{IEEE Transactions on
  Power Systems}, vol.~36, no.~2, pp. 1373--1384, Mar. 2021, conference Name:
  IEEE Transactions on Power Systems.

\bibitem{brockman_openai_2016}
\BIBentryALTinterwordspacing
G.~Brockman, V.~Cheung, L.~Pettersson, J.~Schneider, J.~Schulman, J.~Tang, and
  W.~Zaremba, ``{OpenAI} {Gym},'' \emph{arXiv:1606.01540 [cs]}, Jun. 2016,
  arXiv: 1606.01540. [Online]. Available: \url{http://arxiv.org/abs/1606.01540}
\BIBentrySTDinterwordspacing

\end{thebibliography}

\end{document}